\documentclass[aps,prl, twocolumn, superscriptaddress, longbibliography]{revtex4-2}

\usepackage{times}
\usepackage{amsmath}
\usepackage{amssymb}
\usepackage{graphicx}
\usepackage[caption=false, position=top, singlelinecheck=off, justification=raggedright]{subfig}
\usepackage{color}
\usepackage{pst-node}
\usepackage[unicode]{hyperref}
\hypersetup{unicode=true, colorlinks=true, linkcolor=blue, citecolor=blue}

\usepackage[normalem]{ulem}

\begin{document}

\title{Condensate formation in a dark state of a driven atom-cavity system}

\author{Jim Skulte}\thanks{These authors have contributed equally to this work.}
\affiliation{Zentrum f\"ur Optische Quantentechnologien and Institut f\"ur Laser-Physik, Universit\"at Hamburg, 22761 Hamburg, Germany}
\affiliation{The Hamburg Center for Ultrafast Imaging, Luruper Chaussee 149, 22761 Hamburg, Germany}

\author{Phatthamon Kongkhambut}\thanks{These authors have contributed equally to this work.}
\affiliation{Zentrum f\"ur Optische Quantentechnologien and Institut f\"ur Laser-Physik, Universit\"at Hamburg, 22761 Hamburg, Germany}

\author{Sahana Rao}
\affiliation{Zentrum f\"ur Optische Quantentechnologien and Institut f\"ur Laser-Physik, Universit\"at Hamburg, 22761 Hamburg, Germany}

\author{Ludwig Mathey}
\affiliation{Zentrum f\"ur Optische Quantentechnologien and Institut f\"ur Laser-Physik, Universit\"at Hamburg, 22761 Hamburg, Germany}
\affiliation{The Hamburg Center for Ultrafast Imaging, Luruper Chaussee 149, 22761 Hamburg, Germany}

\author{Hans Ke{\ss}ler}
\affiliation{Zentrum f\"ur Optische Quantentechnologien and Institut f\"ur Laser-Physik, Universit\"at Hamburg, 22761 Hamburg, Germany}

\author{Andreas Hemmerich}
\affiliation{Zentrum f\"ur Optische Quantentechnologien and Institut f\"ur Laser-Physik, Universit\"at Hamburg, 22761 Hamburg, Germany}
\affiliation{The Hamburg Center for Ultrafast Imaging, Luruper Chaussee 149, 22761 Hamburg, Germany}

\author{Jayson G. Cosme}
\affiliation{National Institute of Physics, University of the Philippines, Diliman, Quezon City 1101, Philippines}

\date{\today}

\begin{abstract}   
We demonstrate the formation of a condensate in a dark state of momentum states, in a pumped and shaken cavity-BEC system. The system consists of an ultracold quantum gas in a high-finesse cavity, which is pumped transversely by a phase-modulated laser. This phase-modulated pumping couples the atomic ground state to a superposition of excited momentum states, which decouples from the cavity field. We demonstrate how to achieve condensation in this state, supported by time-of-flight and photon emission measurements. With this, we show that the dark state concept provides a general approach to efficiently prepare complex many-body states in an open quantum system.
\end{abstract}

\maketitle 
While dissipation is in general perceived as a destructive feature of a quantum system, it can also be utilized to engineer nontrivial states, often in conjunction with driving a system out of equilibrium. A prominent experimental platform for this purpose is ultracold quantum gases coupled to high-finesse optical cavities \cite{Baumann2010, Ritsch2013, Klinder2015, Vaidya2018}, due to the well-controlled dissipative channel resulting from the photon emission out of the cavity. Paradigmatic models of light-matter interaction can be explored, such as the celebrated Dicke model that describes the interaction between $N$ two-level atoms with a single quantized light mode \cite{Dicke1954}. The driven-dissipative Dicke model, an extension of the standard Dicke model, captures the scenario, when both external driving and dissipation are present \cite{Kirton2019, Damanet2019}. A wealth of phases, unique to driven light-matter systems, have been proposed and realized using variations of driven Dicke models, such as the three-level Dicke model \cite{Sung1979, Hayn2011, Bastidas2012, Chitra2015, Zhiqiang2017, Soriente2018, Chiacchio2019, Jaksch2019, Stitely2020, Skulte2021, Broers2022, Kongkhambut2021, Lin2022}. In particular, the dissipation channel of the cavity has been utilized to demonstrate the emergence of nonequilibrium or dynamical phases \cite{Habibian2013, Kollath2016, Mivehvar2017, Georges2018, Landini2018, Cosme2019, Dogra2019, Bentsen2019, Holland2020, Kessler2021, Kongkhambut2021, Georges2021, Rodrigo2022, Kongkhambut2022, Dreon2022, Zhang2022}. 

An intriguing class of quantum states in light-matter systems, well known in quantum optics, are the so-called dark states \cite{scullyzubairy1997}. These are superpositions of matter states with relative phases such that the quantum mechanical amplitudes, coupling the different sectors to an irradiated light field, interfere destructively. As a consequence, dark states decouple from the light field. Dark states play a crucial role in physical phenomena, such as stimulated Raman adiabatic passage \cite{Gaubatz1990, Vitanov2017}, electromagnetically induced transparency \cite{Fleischhauer2005, Boller1991}, lasing without inversion \cite{Scully1989, Mompart2000}, and combinations of these topics \cite{Hayn2011, Torre2013, Lin2022, Orioli2022}. In conventional quantum optics scenarios, dark states typically arise on a single particle level. In this Letter, we use the dark state concept in a many-body context, specifically condensation. Our study suggests how the concept of dark state formation can be utilized in the context of quantum state engineering via dissipation. 

\begin{figure}[!htpb]
\centering
\includegraphics[width=1\columnwidth]{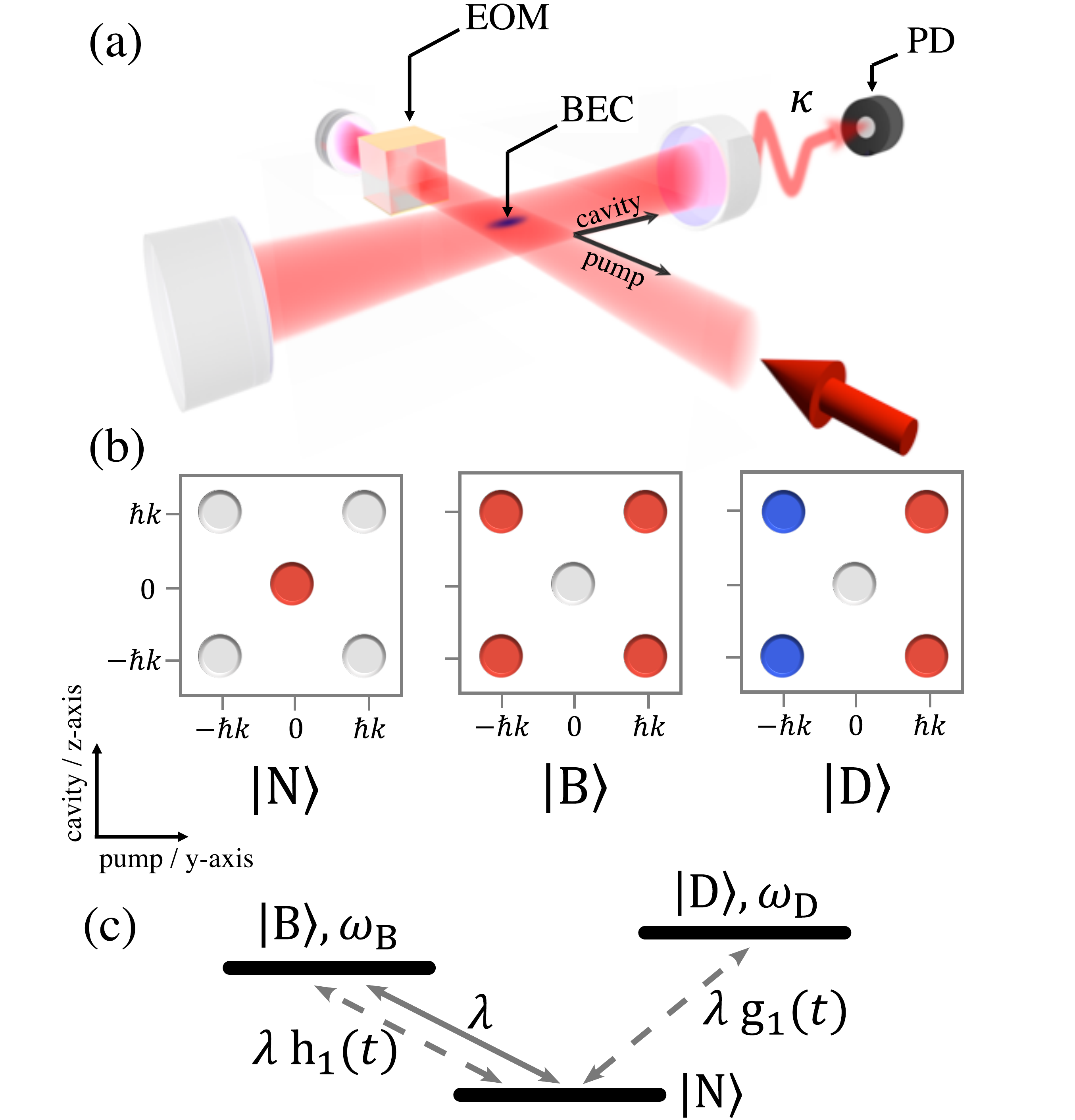}
\caption{(a) Sketch of the experimental setup. An electro-optical modulator (EOM) is used to modulate the phase of the pump field, which results in shaking the pump potential. (b) Sketch of the momentum distribution of the three relevant superpositions of momentum modes, the normal state $|\mathrm{N}\rangle$, the bright state $|\mathrm{B}\rangle$ and the dark state $|\mathrm{D}\rangle$, which form a three-level Dicke model shown in (c) with the atom-cavity coupling $\lambda$ and the shaking-induced functions $h_1(t)$ and $g_1(t)$. The colors in (b) represent the phase of the momentum states, where blue indicates a phase shift of $\pi$ relative to red.} 
\label{fig:1} 
\end{figure}

In this Letter, we demonstrate in theory and experiment a robust condensate formation in a dark state of a driven atom-cavity system, approximately described by a parametrically driven three-level open Dicke model introduced in Refs.~\cite{Skulte2021, Kongkhambut2021}. We consider a Bose-Einstein condensate (BEC) prepared in a high-finesse cavity, which is transversely pumped with a shaken one-dimensional optical lattice, as sketched in Fig.~\ref{fig:1}(a). Previously, we explored the weakly resonantly driven scenario leading to an incommensurate time crystal (ITC) \cite{Cosme2019, Skulte2021, Kongkhambut2021}. Here, technical improvements in our setup allowed us to study theoretically and experimentally the so far unexplored regime of strong driving and a wider range of driving frequencies, which reveals that the ITC has transient character in certain parameter regimes, such that the atoms relax into a dark state of the atom-cavity system eventually. 

To understand the dark state and to identify the relevant driving parameters, we employ the time-dependent atom-cavity Hamiltonian in Refs.~\cite{suppmat, Skulte2021} and an approximative parametrically driven three-level Dicke model \cite{Skulte2021, Kongkhambut2021}, which includes only three atomic modes denoted as $|\mathrm{N}\rangle$, $|\mathrm{B}\rangle$, and $|\mathrm{D}\rangle$, in a plane-wave expansion of the atomic field operator. These modes are illustrated in terms of their momentum components in Fig.~\ref{fig:1}(b) and form the V-shaped three-level system sketched in Fig.~\ref{fig:1}(c). The normal state $|\mathrm{N} \rangle \equiv |(0,0) \hbar k \rangle$ corresponds to a homogeneous density in real space, wherein all atoms occupy the lowest momentum mode $\{p_y, p_z\} = \{0,0\} \hbar k$ ($k$ is the wave number of the pump field). The pump leads to a light shift of $-\epsilon_\mathrm{p}\omega_\mathrm{rec}/2$, where $\epsilon_\mathrm{p}$ is the unitless pump intensity and $\omega_\mathrm{rec}$ is the atomic recoil frequency. The bright state $|\mathrm{B}\rangle \equiv$ $\sum_{\nu,\mu\in\{-1,1\}} |\nu \hbar k,\mu \hbar k\rangle$ is defined as the in-phase superposition of the $\{\pm1,\pm1\}\hbar k$ momentum modes as depicted in Fig.~\ref{fig:1}(b). The real-space wave function of this state is $\propto \cos(ky)\cos(kz)$, which has even parity with respect to the inversion $(y,z)\rightarrow (-y,-z)$. It exhibits a kinetic energy of $2 E_\mathrm{rec}$ and is light shifted by the pump wave by $-3\epsilon_\mathrm{p}\omega_\mathrm{rec}/4$ such that its frequency separation relative to $|\mathrm{N}\rangle$ is $\omega_{\mathrm{B}} = (2 - \epsilon_\mathrm{p}/4)\, \omega_{\mathrm{rec}}$. The dark state $|\mathrm{D}\rangle \equiv$ $\sum_{\nu,\mu\in\{-1,1\}} \nu |\nu \hbar k,\mu \hbar k\rangle$ is defined as the out-of-phase superposition of the $\{+1,\pm 1\}\hbar k$ and  $\{-1,\pm 1\}\hbar k$ momentum modes. In real space, its order parameter is $\propto \sin(ky)\cos(kz)$, which has odd parity under the inversion $(y,z)\rightarrow (-y,-z)$. 

The density distributions of the dark state $|\mathrm{D}\rangle$ and the bright state $|\mathrm{B}\rangle$ both prohibit collective scattering of photons into the cavity. Nonetheless, any admixture of the normal state $|\mathrm{N}\rangle$ to the bright state $|\mathrm{B}\rangle$ leads to a checkerboard pattern of the atomic density that allows pump photons to scatter into the cavity, which is the reason why we refer to $|\mathrm{B}\rangle$ as a bright state. Above a critical pump strength, the system forms a superradiant (SR) phase as its stationary state, in which a superposition of $|\mathrm{B}\rangle$ and $|\mathrm{N}\rangle$ produces a density grating trapped by the intracavity optical lattice composed of the pump and cavity fields. In contrast to $|\mathrm{B}\rangle$, the density grating of the dark state $|\mathrm{D}\rangle$, due to its odd parity is shifted along the pump direction by a quarter of the pump wavelength, such that the atomic positions coincide with the nodes of the pump lattice, motivating our terminology of bond-density waves in Refs.~\cite{Skulte2021, Kongkhambut2021}. Hence, even if $|\mathrm{N}\rangle$ is admixed to the dark state $|\mathrm{D}\rangle$, scattering of pump photons remains suppressed, meaning that for any superposition of the normal and the dark state collective scattering of photons into the cavity cannot occur. The dark state $|\mathrm{D}\rangle$ exhibits the same kinetic energy $2 E_\mathrm{rec}$ as $|\mathrm{B}\rangle$, while its light shift due to the pump lattice is only $-\epsilon_\mathrm{p}\,\omega_\mathrm{rec}/4$. Thus, its frequency relative to that of $|\mathrm{N}\rangle$ is $\omega_{\mathrm{D}} = (2 + \epsilon_\mathrm{p}/4)\,\omega_{\mathrm{rec}}$. 

\begin{figure}[!htpb]
\centering
\includegraphics[width=1.0\columnwidth]{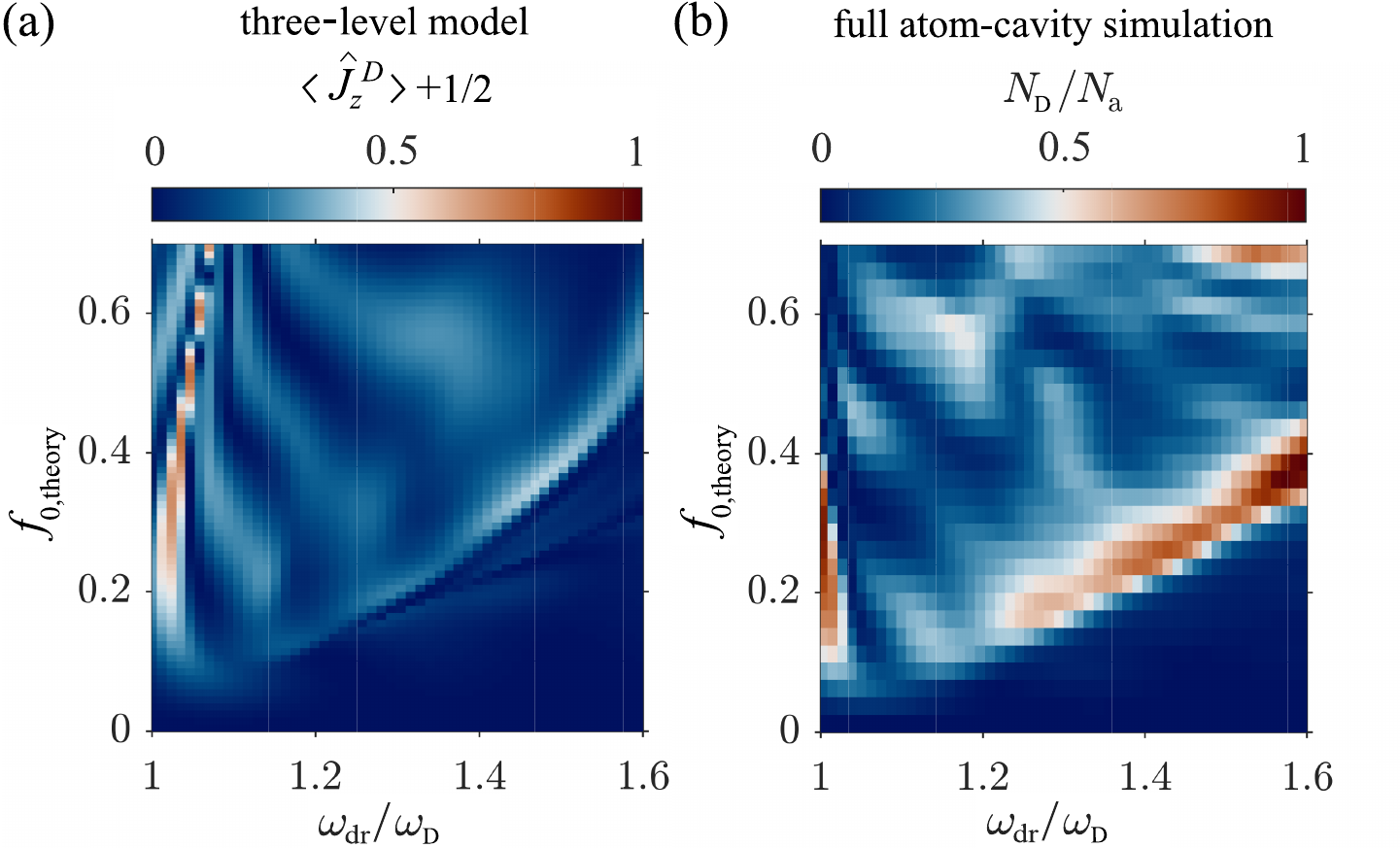}
\caption{(a),(b) Population of the dark state for different driving frequencies $\omega_\mathrm{dr}$ and driving strengths $f_{0,\mathrm{theory}}$. The driving frequency axis is rescaled by the characteristic frequency of the dark state, $\omega_\mathrm{D}$. (a) the results from the three-level model and (b) the full atom-cavity simulation. The phase diagrams are constructed for 7 driving cycles.}
\label{fig:2} 
\end{figure}

To excite the dark state, we shake the pump lattice by introducing a time-dependent phase in the pump field, $\cos(ky+\phi(t))$, where $\phi(t) = f_0 \sin (\omega_\mathrm{dr} t)$, $f_0$ is the driving strength, and $\omega_\mathrm{dr}$ is the driving frequency. The excitation mechanism is readily understood by means of the three-level Dicke model $\hat{H}=\hat{H}_\mathrm{stat}+\hat{H}_\mathrm{dyn}$ with a static part
\begin{align}
\label{eq:ham1}
\hat{H}_\mathrm{stat}/ \hbar &= \omega \hat{a}^\dagger \hat{a} +\left[\omega_{\mathrm{B}}-\Omega(f_0)\right]\hat{J}^{\mathrm{B}}_z   + \left[\omega_{\mathrm{D}}+\Omega(f_0)\right]\hat{J}^{\mathrm{D}}_z  \nonumber \\
&+\frac{2\lambda}{\sqrt{N_{\mathrm{a}}}}\left(\hat{a}^\dagger+\hat{a} \right) J_0(f_0)\hat{J}^{\mathrm{B}}_x
\end{align}
and a dynamical part
\begin{align}
\label{eq:ham2}
\hat{H}_\mathrm{dyn}/ \hbar &=h_2(t)\Delta\omega_\mathrm{BD}\left(\hat{J}^{\mathrm{D}}_z -\hat{J}^{\mathrm{B}}_z \right)+2g_2(t) \Delta\omega_\mathrm{BD}\hat{J}^{\mathrm{BD}}_x  \nonumber  \\ 
&+\frac{4\lambda}{\sqrt{N_{\mathrm{a}}}}\left(\hat{a}^\dagger+\hat{a} \right) \left(h_1(t)\hat{J}^{\mathrm{B}}_x -g_1(t) \hat{J}^{\mathrm{D}}_x \right),
\end{align}
where $\Omega(f_0)=\frac{\epsilon_\mathrm{p}\omega_\mathrm{rec}}{4}(1-J_0(2f_0))$, $\Delta\omega_\mathrm{BD}=\left(\omega_{\mathrm{B}}-\omega_{\mathrm{D}} \right) $, $h_m(t)=\sum_{n=1}^\infty J_{2n}(m f_0)\cos(2n\omega_\mathrm{dr} t)$, $g_m(t)=\sum_{n=1}^\infty J_{2n-1}(mf_0)\sin((2n-1)\omega_\mathrm{dr} t)$, and $J_n(r)$ is the $n^\mathrm{th}$ Bessel function of the first kind. The time-dependent terms introduced by the pump lattice shaking are $h_m(t)$ and $g_m(t)$. Details on the derivation of this Hamiltonian are given in the Supplemental Material \cite{suppmat}. The pseudospin operators $\hat{J}^\mathrm{B}_\mu$ ($\mu \in \{x,y,z\}$) describe the coupling to the bright state since $\hat{J}^\mathrm{B}_+ \equiv \hat{J}^\mathrm{B}_x + i \hat{J}^\mathrm{B}_y$ $= |\mathrm{B}\rangle\langle \mathrm{N}|$. Accordingly, $\hat{J}^\mathrm{D}_\mu$ is related to the dark state as $\hat{J}^\mathrm{D}_+ \equiv \hat{J}^\mathrm{D}_x + i \hat{J}^\mathrm{D}_y$ $= |\mathrm{D}\rangle\langle \mathrm{N}|$. We see from the last term of Eq.~\eqref{eq:ham2} that $|\mathrm{D}\rangle$ can be coupled to the cavity mode via the time-dependent shaking of the pump, resulting in a periodic coupling between $|\mathrm{D}\rangle$ and $|\mathrm{N}\rangle$. Note that the necessary nonzero amplitude $g_1(t)$ can be provided by phase modulation, which breaks the discrete translation symmetry along the pump axis, but not by amplitude modulation. We consider the recoil-resolved regime, i.e., the loss rate of the cavity photons $\kappa$ is comparable to the recoil frequency $\omega_\mathrm{rec}$, which for our system is $\omega_\mathrm{rec}= 2 \pi \times 3.6~\mathrm{kHz}$. We emphasize the importance of this regime \cite{Kessler2014, Klinder2016} to protect the dark state from detrimental resonant excitations to higher energy momentum states.

Next, we discuss the dynamics of the system by solving the semiclassical equations of motion of the three-level model Eq.~\eqref{eq:ham2} and those of the atom-cavity Hamiltonian \cite{suppmat} including fluctuations due to photon emission out of the cavity. For the three-level model, the dark state occupation is $\langle \hat{J}^\mathrm{D} \rangle + 1/2$. For the full atom-cavity model, we apply the following protocol: the pump laser strength is linearly increased within $10~\mathrm{ms}$, such that we always initially prepare the SR phase. After a holding time of $0.5~\mathrm{ms}$, the phase of the pump lattice is modulated for 7 driving cycles, starting at $t=t_0$. We choose 7 driving cycles since, as is later seen in the experiment, the dark state occupation $N_\mathrm{D}$ is found to equilibrate after 6 driving cycles due to heating \cite{suppmat}. Subsequently, we adiabatically ramp-down the pump strength in $0.5\,$ms and calculate $N_\mathrm{D}$ as the sum of the occupations in the $\{\pm 1,\pm 1\}\hbar k$ modes. The ramp down is necessary to remove all $\{\pm 1,\pm 1\}\hbar k$ populations, associated with $|\mathrm{B}\rangle$ rather than $|\mathrm{D}\rangle$, by transferring $|\mathrm{B}\rangle$ into $|\mathrm{N}\rangle$, which does not affect $|\mathrm{D}\rangle$. In Fig.~\ref{fig:2}, we construct the phase diagrams of the three-level and the full models, plotting $N_\mathrm{D}$ for different driving parameters. Our previous work on the emergence of an ITC involved the regime around $\omega_\mathrm{dr} \in [1,1.2]\times \omega_\mathrm{D}$ and $f_\mathrm{0,theory}<0.4$ \cite{Skulte2021, Kongkhambut2021}. We find qualitative agreement between the numerical simulations of the full atom-cavity system and the driven three-level Dicke model as seen in Fig.~\ref{fig:2}. Significant occupation of the dark state is observed in a large area of the phase diagram for $\omega_\mathrm{dr} > \omega_\mathrm{D}$ and also in a small area close to the resonance $\omega_\mathrm{dr} \approx \omega_\mathrm{D}$. We note that the area in the driving parameter space, where the dark state becomes dynamically occupied, is larger in the full atom-cavity model as compared to the three-level Dicke model. This can be attributed to the $\{0,\pm 2\}\hbar k$ and $\{\pm 2,0\}\hbar k$ modes, which are neglected in the three-level model \cite{suppmat}. Atoms in these modes may be transferred to the dark state upon scattering photons into the cavity, thus increasing its efficient population. This process competes with a direct resonant transfer of atoms into the second band of the pump wave without scattering photons into the cavity, which impedes efficient population of the dark state as detailed in the Supplemental Material \cite{suppmat}. The respective resonance frequency arises in Fig.~\ref{fig:2} for $\omega_\mathrm{dr}/ \omega_\mathrm{D} \approx 1.7$, i.e. slightly outside the shown range.

\begin{figure}[!htbp]
\centering
\includegraphics[width=1\columnwidth]{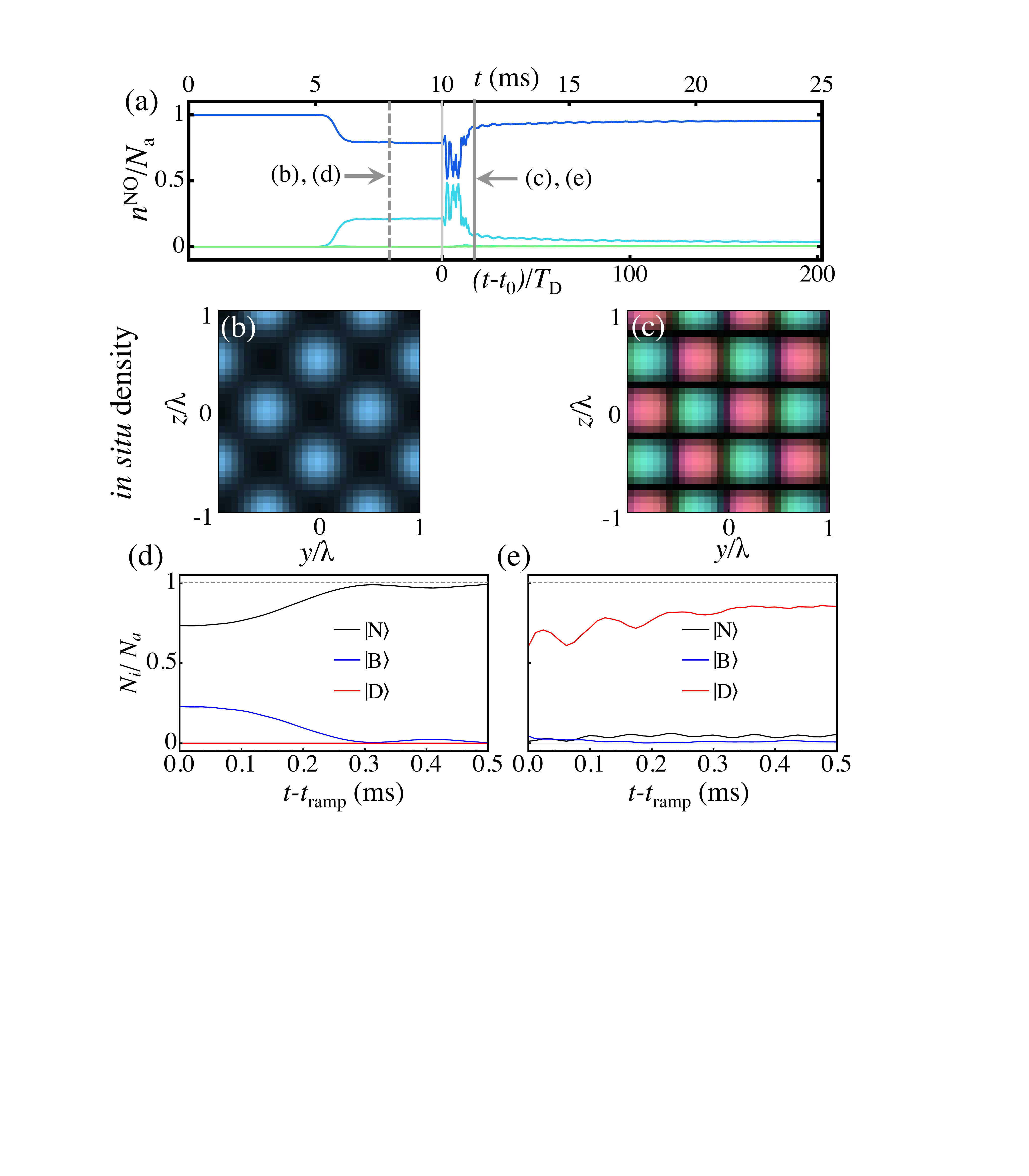}
\caption{(a) Simulations of  the evolution of the  three highest eigenvalues of the single-particle correlation function are shown. Gray dashed and solid vertical lines denote, respectively, the times when the snapshots of the single-particle densities in (b) and (c) are taken. The real-space densities in (b) and (c) are color coded to show the phase of $\Psi(y,z)$. (d), (e) Evolution of the occupations of $|\mathrm{N}\rangle$, $|\mathrm{B}\rangle$, and $|\mathrm{D}\rangle$, while the pump is adiabatically ramped down. Panels (d) and (e), respectively, correspond to initial conditions according to the dashed and solid gray vertical lines in (a).}
\label{fig:3} 
\end{figure} 

\begin{figure*}[!htpb]
\centering
\includegraphics[width=2\columnwidth]{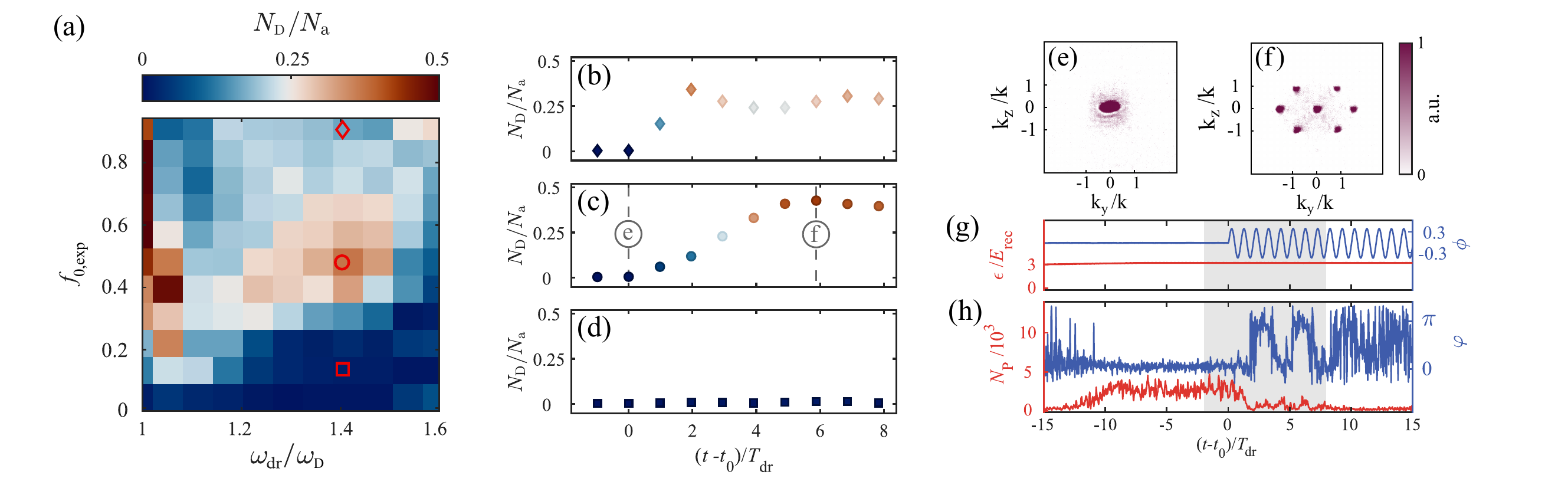}
\caption{Experimental phase diagram of the population of the dark state for different driving frequencies $\omega_\mathrm{dr}$ and driving strengths $f_{0,\mathrm{exp}}$. The driving frequency axis is rescaled by the characteristic frequency of the dark state, $\omega_\mathrm{D}$. The phase diagram is constructed for 6 driving cycles. (b)-(d) Population of the dark state as a function of the driving cycles for the parameter sets marked by a diamond, a circle and a rectangle in (a). The population of the dark state is rescaled by the total particle number $N_\mathrm{a}$ for different driving cycles derived from time-of-flight (TOF) images in (b)-(d). Examples of TOF images are provided before shaking starts at $t=0$ (e) and after around 6 driving cycles (f). All TOF images are obtained after an adiabatic ramp-down of the pump wave and ballistic expansion of 25ms. (g) Time sequence for the pump strength (red) and the phase $\phi$ of the pump field (blue). (h) Phase difference $\varphi$ between the pump and intracavity field (blue) and intracavity photon number $N_\mathrm{P}$ (red) for the driving parameters marked by a circle in (a).}
\label{fig:4} 
\end{figure*}  
Next, we employ the truncated Wigner approximation (TWA) to capture the leading-order quantum effects \cite{Polkovnikov2010, Cosme2019, Hans2019, Tuquero2022}. We include not only the dissipation due to photon emission out of the cavity but also the associated fluctuations. We further demonstrate that the observed dark state is indeed a finite momentum condensate by calculating the eigenvalues of the single-particle correlation function at equal time, $\langle \Psi(y,z)^\dagger\Psi(y',z') \rangle$ for our full atom-cavity model. This appears in the Penrose-Onsager criterion for condensates and its largest eigenvalue corresponds to the condensate fraction \cite{Penrose1956}. We denote the eigenvalues as $n_\mathrm{NO}$. We show in Fig.~\ref{fig:3}(a) the $n_\mathrm{NO}$ obtained from TWA simulations for the same pump protocol used in Fig.~\ref{fig:2}(b), but without the final ramp-down of the pump wave. When the system enters the SR phase (at about $5.2\,$ms), the condensate fragments manifested in the reduction of the largest eigenvalue and the corresponding increase of the second largest eigenvalue \cite{Lode2017}. The real-space density $|\Psi(y,z)|^2$ shown in Fig.~\ref{fig:3}(b) illustrates the prevailing SR phase at the time indicated by the dashed gray line, before driving starts at $t=t_0$. In Fig.~\ref{fig:3}(c), we show $|\Psi(y,z)|^2$ at the time indicated by the solid gray line, after driving has acted for about $0.6\,$ms, indicating a substantial population of the dark state. The zeros (black regions) coincide with the intensity maxima of the pump lattice along the $y$-direction, while there is no significant standing wave potential along the cavity direction. The different colors in Fig.~\ref{fig:3}(c) denote opposite phases of $\Psi(y,z)$. In Figs.~\ref{fig:3}(d) and \ref{fig:3}(e), we show the occupations of the relevant states as the pump lattice is ramped down at the times indicated by the dashed (d) and solid (e) gray lines. It can be seen in Fig.~\ref{fig:3}(d) that for the SR phase (prevailing at the time denoted by the dashed vertical line in Fig.~\ref{fig:3}(a)) practically all atoms are transferred back to the normal state $|\mathrm{N}\rangle$ after the ramp-down. On the other hand, for the driven case in Fig.~\ref{fig:3}(e), associated with the time indicated by the solid gray line in (a), the dark state $|\mathrm{D}\rangle$ has the largest occupation at $t=t_\mathrm{ramp}$. After the ramp-down, its occupation increases further, forming a long-lived state, compared to the decay time of the SR state. These results corroborate that the population of the $\{\pm1,\pm1\}\hbar k$ modes after the pump is adiabatically switched off, is the appropriate observable to quantify the {\it in-situ} occupation of the dark state.

Finally, we experimentally demonstrate driving-induced condensation into a dark state of the atom-cavity system \cite{suppmat}. We present in Fig.~\ref{fig:4}(a) the resulting experimental phase diagram of the occupation of $N_{\mathrm{D}}$ for varying driving parameters. We find qualitative agreement with the theoretical phase diagrams depicted in Fig.~\ref{fig:2}. Due to technical reasons, such as atom losses, a complete population inversion into the dark state, as seen in the numerical simulations, is not observed in the experiment. We note that there is a slight difference between the numerical and the experimental results for the driving strength needed to populate the dark state.  This can possibly be attributed to the pump in the experiment having a nonzero width in frequency space, so that the effective pump power is smaller than it would be for monochromatic pump beam. Therefore the experimental realization might require a nominally larger pump power than in the theoretical model.

Figs.~\ref{fig:4}(b)-(d) show the occupation of the dark state for varying numbers of driving cycles and fixed driving frequencies. Each panel corresponds to a value of the driving strength $f_\mathrm{0,exp}$ indicated by the red markers in Fig.~\ref{fig:4}(a). Between the red circular and the red rectangular marker, there is a sharp transition from large occupation of $|\mathrm{D}\rangle$ (see also Fig.~\ref{fig:4}(c)) towards a region where $|\mathrm{D}\rangle$ is practically unoccupied (see also Fig.~\ref{fig:4}(d)). In the limit of strong driving around the diamond-shaped marker in Fig.~\ref{fig:4}(a), the dark state becomes highly occupied after only 2 driving cycles, but the occupation number slightly decreases again for larger numbers of driving cycles as shown in Fig.~\ref{fig:4}(b). This is explained by the excitation of the $|\pm 2 \hbar k,0 \rangle$ modes, as discussed below. Each data point is obtained via averaging over $10$ TOF images. We also present the corresponding TOF images (see Figs.~\ref{fig:4}(e) and \ref{fig:4}(f)) at two instances of time, i.e., at $t= t_0 $ before driving is started, and after six driving cycles at $t = t_0 + 6\, T_{\mathrm{dr}}$ as indicated in Fig.~\ref{fig:4}(c). These TOF images correspond to the spatial orders calculated in Figs.~\ref{fig:3}(d) and \ref{fig:3}(e). We display the time evolution of the cavity field for a single experimental realization in Fig.~\ref{fig:4}(h) showcasing the vanishing intracavity light field as a macroscopic fraction of the atoms occupy the dark state.

For the case depicted in Fig.~\ref{fig:4}(c), we find that initially, $N_{\mathrm{D}}$ increases and saturates beyond 6 driving cycles. The system approaches a steady state because of atom losses before all atoms can be transferred into the dark state. In contrast to the SR phase in Fig.~\ref{fig:4}(e), the large occupation of the four momentum components $\{\pm1,\pm1\}\hbar k$ in Fig.~\ref{fig:4}(f) in combination with the small intracavity photon number in Fig.~\ref{fig:4}(h) indicates a large occupation of the dark state $|\mathrm{D}\rangle$. Furthermore, a substantial fraction of atoms populates the $\{\pm2,0\}\hbar k$ momentum modes as the driving frequency is tuned close to the resonance frequency for excitation to the second band of the pump wave. This process inhibits efficient population of the dark state as is discussed in the Supplemental Material \cite{suppmat}. For reasons explained in Ref. \cite{suppmat}, in the experiment, the respective resonance is shifted to $\omega_\mathrm{dr}/ \omega_\mathrm{D} \approx 1.45$, i.e., within the domain shown in Fig.~\ref{fig:4}(a), acting to suppress the dark state population on the right side of the red circle.

In conclusion, in an atom-cavity system pumped by a periodically shaken standing wave, we have found that in a specific parameter domain, a stationary excited dark state condensate emerges, in which scattering of pump photons into the cavity mode is suppressed. We show that a three-level Dicke model captures this phenomenon qualitatively. Both theoretically and experimentally, we observe that, upon adiabatic ramp-down of the pump wave, the atomic condensate in the dark state is essentially unaffected, while the bright sector of the system undergoes a dynamical phase transition \cite{Klinder2015}. Our work points out a general approach to form stationary excited many-body states using the concept of dark states known from single-particle quantum optics.

\begin{acknowledgments}
We thank C. Georges, J. Klinder, and L. Broers for helpful discussions. This work was funded by the UP System Balik PhD Program (OVPAA-BPhD-2021-04), the QuantERA II Programme that has received funding from the European Union's Horizon 2020 research and innovation programme under Grant Agreement No 101017733, the Deutsche Forschungsgemeinschaft (DFG, German Research Foundation) ``SFB-925" Project No 170620586 and the Cluster of Excellence ``Advanced Imaging of Matter" (EXC 2056), Project No. 390715994. J.S. acknowledges support from the German Academic Scholarship Foundation. 
\end{acknowledgments}

\nocite{abramowitz,Keeling2010}

\bibliography{references_dark_state}

\end{document}


\title{Supplemental Material for \\ Condensate formation in a dark state of a driven atom-cavity system}

\author{Jim Skulte}\thanks{These authors have contributed equally to this work.}
\affiliation{Zentrum f\"ur Optische Quantentechnologien and Institut f\"ur Laser-Physik, Universit\"at Hamburg, 22761 Hamburg, Germany}
\affiliation{The Hamburg Center for Ultrafast Imaging, Luruper Chaussee 149, 22761 Hamburg, Germany}

\author{Phatthamon Kongkhambut}\thanks{These authors have contributed equally to this work.}
\affiliation{Zentrum f\"ur Optische Quantentechnologien and Institut f\"ur Laser-Physik, Universit\"at Hamburg, 22761 Hamburg, Germany}

\author{Sahana Rao}
\affiliation{Zentrum f\"ur Optische Quantentechnologien and Institut f\"ur Laser-Physik, Universit\"at Hamburg, 22761 Hamburg, Germany}

\author{Ludwig Mathey}
\affiliation{Zentrum f\"ur Optische Quantentechnologien and Institut f\"ur Laser-Physik, Universit\"at Hamburg, 22761 Hamburg, Germany}
\affiliation{The Hamburg Center for Ultrafast Imaging, Luruper Chaussee 149, 22761 Hamburg, Germany}

\author{Hans Ke{\ss}ler}
\affiliation{Zentrum f\"ur Optische Quantentechnologien and Institut f\"ur Laser-Physik, Universit\"at Hamburg, 22761 Hamburg, Germany}

\author{Andreas Hemmerich}
\affiliation{Zentrum f\"ur Optische Quantentechnologien and Institut f\"ur Laser-Physik, Universit\"at Hamburg, 22761 Hamburg, Germany}
\affiliation{The Hamburg Center for Ultrafast Imaging, Luruper Chaussee 149, 22761 Hamburg, Germany}

\author{Jayson G. Cosme}
\affiliation{National Institute of Physics, University of the Philippines, Diliman, Quezon City 1101, Philippines}
\date{\today}

\maketitle

\section{BEC, cavity, and pump beam properties}
The experimental setup, as sketched in Fig.~1(a) in the main text, is comprised of a magnetically trapped BEC of $N_\mathrm{a} = 4\times 10^4$  $^{87}$Rb atoms, dispersively coupled to a fundamental mode of a narrowband high-finesse optical cavity. The trap creates a harmonic potential with trap frequencies $(\omega_x, \omega_y,\omega_z) = 2\pi \times (119.0,102.7,24.7)$~Hz. The corresponding Thomas-Fermi radii of the ensemble are $(r_x,r_y,r_z) = (3.7, 4.3, 18.1)~\mu m$. These radii are significantly smaller than the size of the Gaussian shaped pump beam, which has a waist of $w_\mathrm{pump}~\approx~125~\mu m$. The pump beam is oriented transversally, with respect to the cavity axis, and retro-reflected to form a standing wave. It passes through an electro-optic modulator (EOM) twice. An AC voltage is applied to the EOM to modulate the phase of the pump field, which leads to an effective shaking of the pump lattice potential. 

The pump laser is stabilized to the cavity resonance using high bandwidth servo electronics. As a drawback, the pump light is not strictly monochromatic and besides the narrow carrier, the spectrum contains two servo bumps with a frequency shift of roughly $\pm 2$~MHz. We estimate the light power with in these side peaks being about $30\%$ of the total light power. Since this light is far detuned, with respect to the cavity resonance, it cannot contribute to scatter photons into the cavity. In contrast, light of all frequencies contribute to the depth of the standing wave potential, and hence, contributes to the shift of the resonance frequency of the dark state $\omega_\mathrm{D} = (2 + \epsilon_\mathrm{p}/4)\,\omega_\mathrm{rec}$. Therefore the dark state resonance frequency in the experiment is larger than the one used in our theoretical models.

The cavity field has a decay rate of $\kappa \approx ~2\pi \times3.6\,$kHz, which equals the recoil frequency $\omega_\mathrm{rec} = E_\mathrm{rec}/\hbar  =~2\pi \times3.6\,$kHz for $^{87}$Rb atoms at the pump wavelength of $\lambda_\mathrm{P} = 803.00\,$nm. The pump laser is red detuned with respect to the relevant atomic transition of $^{87}$Rb at 794.98~nm. The maximum light shift per atom is $U_0 = 2\pi \times 0.4~\mathrm{Hz}$.

\section{Cavity field detection}
Our experimental system is equipped with two detection setups for the light leaking out of the cavity. On one side of the cavity, we use a single photon counting module (SPCM), which gives access to the intensity of the intracavity field and the associated photon statistics. On the other side of the cavity, a balanced heterodyne detection setup is installed, which uses the pump beam as a local reference. The beating signal of the local oscillator with the light leaking out of the cavity allows for the observation of the time evolution of the intracavity photon number $N_\mathrm{P}$ and the phase difference between the pump and the cavity field $\varphi$.

\section{Experimental protocol to obtain the population of the dark state $N_{\mathrm{D}}$}
To obtain the population of the dark state $N_{\mathrm{D}}$ experimentally, we ramp down the pump laser strength adiabatically within 0.5~ms, similar to the theoretical protocol described in the context of Fig.2(b) in the main text. Subsequently, a ballistic expansion of $25\,$ms is applied and an absorption image of the resulting density distribution is recorded, time-of-flight (TOF). Finally, $N_\mathrm{D}$ is obtained by summing up the occupations around the momentum modes $ \{\pm 1,\pm 1\}\hbar k$, in accordance with the findings in Figs.3(d) and Fig.3(e) in the main text.

\begin{figure}[!htpb]
\centering
\includegraphics[width=0.65\columnwidth]{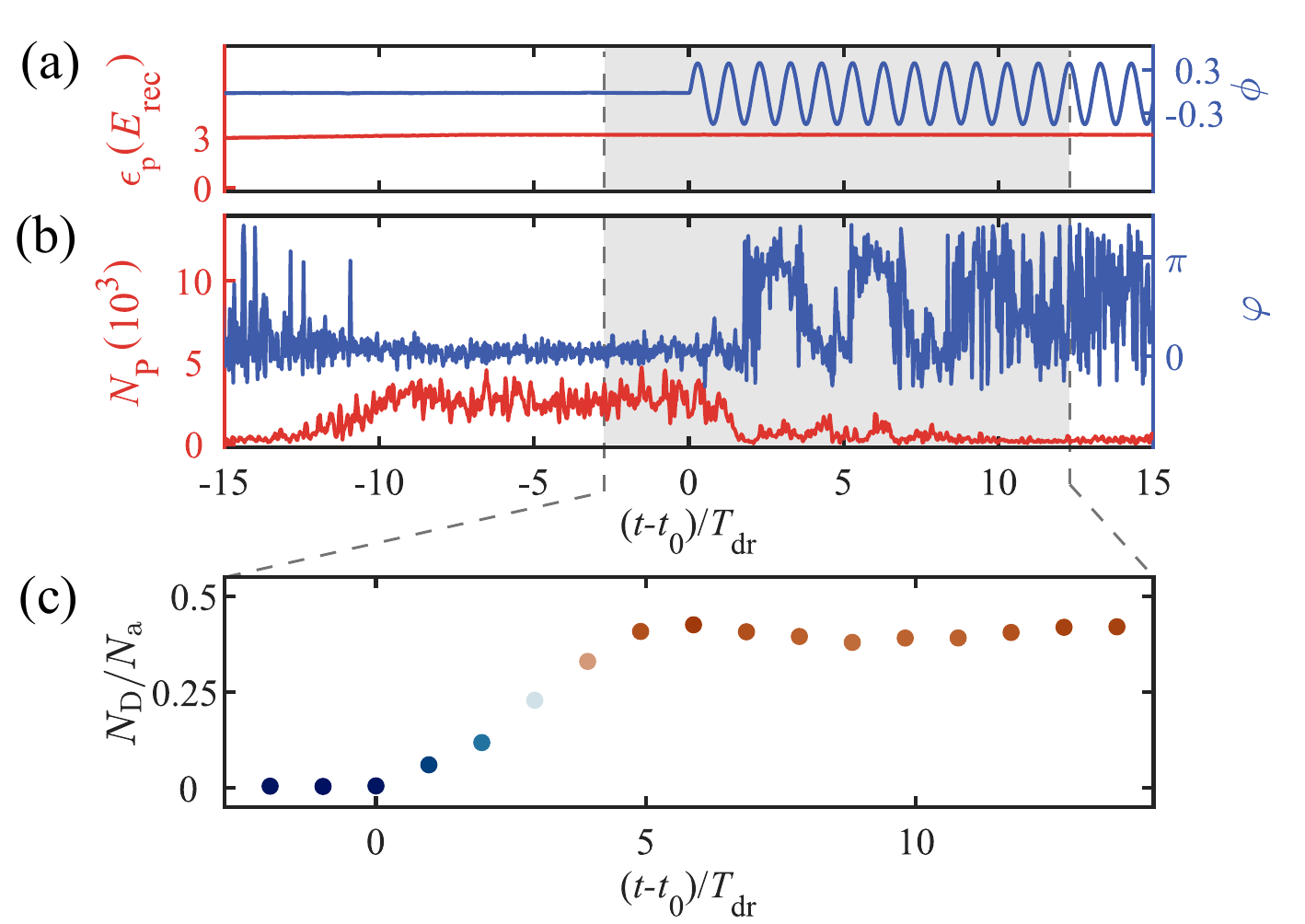}
\caption{Typical experimental run. (a) Experimental protocol for the intensity (red) and the phase (dashed blue) of the pump field. (b) In situ number of photons (red) in the cavity $N_\mathrm{P}$ and the corresponding phase difference $\varphi$ (blue) between the pump and cavity fields. (c) Population of the dark state rescaled by the total particle number $N_\mathrm{a}$ (coherent fraction) for different number of driving cycles derived from TOF images.}
\label{sf:1} 
\end{figure}

\section{Dynamics during a typical experimental run}

\begin{figure}[!htbp]
\centering
\includegraphics[width=1\columnwidth]{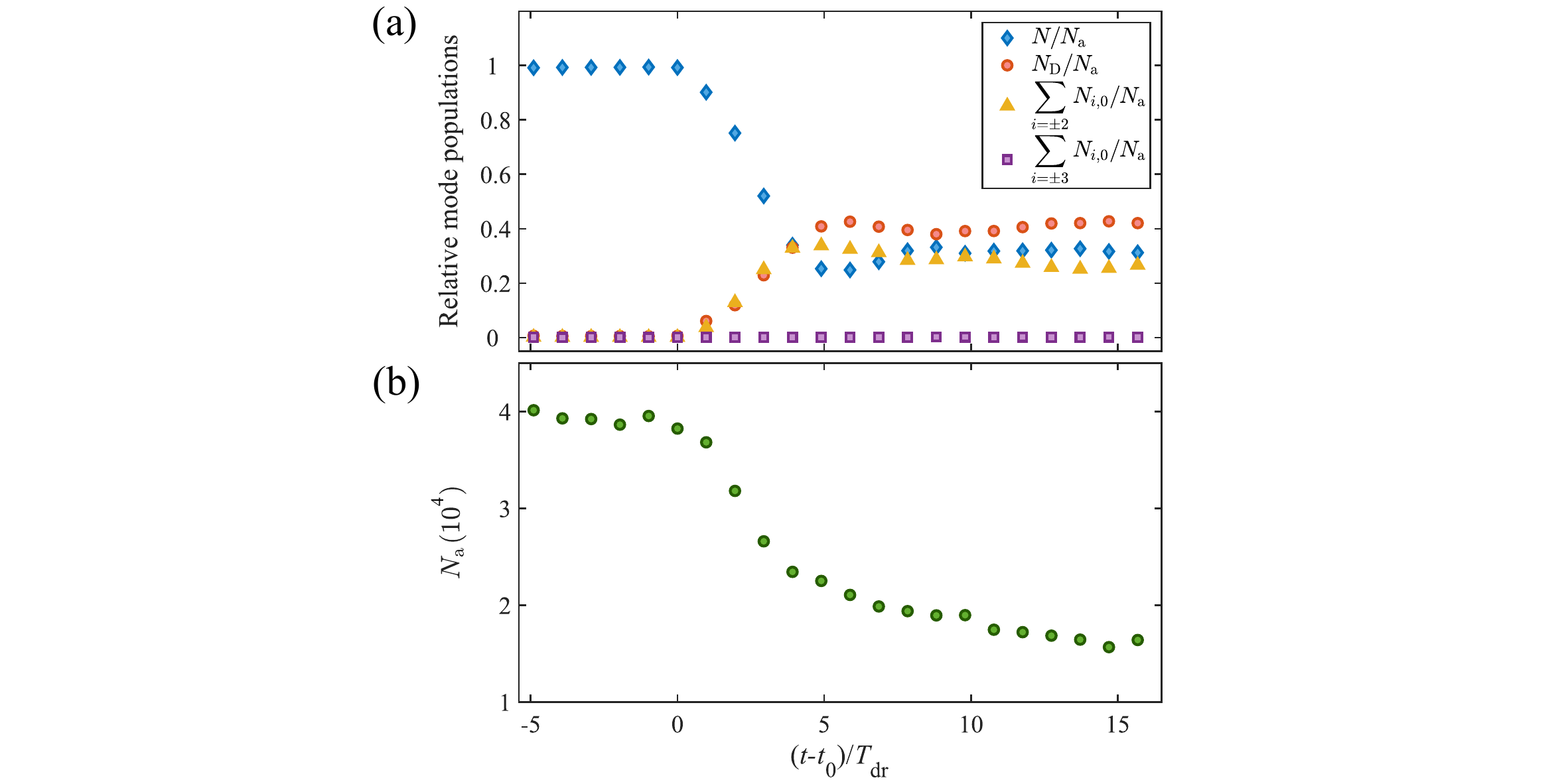}
\caption{(a) Time evolution of the population of the relevant momentum modes normalized to the total number of atoms (coherent fraction). (b) Time evolution of the total number of coherent atoms $N_\mathrm{a}$.}
\label{sf:2} 
\end{figure}

Our experimental sequence starts by preparing a BEC and overlap it with the $\mathrm{TEM}_{00}$ mode of our cavity. We linearly increase the pump strength $\epsilon$ to its desired value to initialise the system into the self-organized superradiant phase. This is indicated by a finite photon number $N_\mathrm{P}$ (red trace in Fig.~\ref{sf:1}(b)) and by the fixed phase difference $\varphi$ between the pump and cavity fields. At time $t - t_\mathrm{0}=0$ we switch on the phase  modulation of the pump lattice, which leads to a periodic shaking of the optical potential. The system starts to oscillate between the two possible self-organized density patterns, which can be seen by the phase difference $\varphi$ switching between 0 and $\pi$. This is accompanied by an increase in the population of the dark state $N_\mathrm{D}/N_\mathrm{a}$ until it reaches its maximum value at time $t-t_0=5~T_\mathrm{dr}$,  where $T_\mathrm{dr}$ is the driving period. Due to the increasing population of the dark state the atoms step by step decouple from the cavity field and slowly stop scattering photons from the pump into the cavity and vice versa. The photon number $N_\mathrm{P}$ approaches zero and the light field phase $\varphi$ shows random values between 0 and $2\pi$. The system is now in a steady state and the population of the dark state, normalized to the total number of coherent atoms $N_\mathrm{D}/N_\mathrm{a}$, stays constant. Fig.~\ref{sf:2}(a) shows the dynamics of the relative population of all relevant momentum modes. Fig.~\ref{sf:2}(b) depicts the corresponding time evolution of $N_\mathrm{a}$. As soon as the shaking starts ($t - t_0=0$), the total particle number $N_\mathrm{a}$ drops rapidly due to cavity-field-induced heating. After the atoms are decoupled from the cavity field, the heating rate decreases.

\section{Comparison of the relative population of the dark state for pump light close and far detuned with respect to the cavity resonance}

\begin{figure}[htbp]
\centering
\includegraphics[width=0.6\columnwidth]{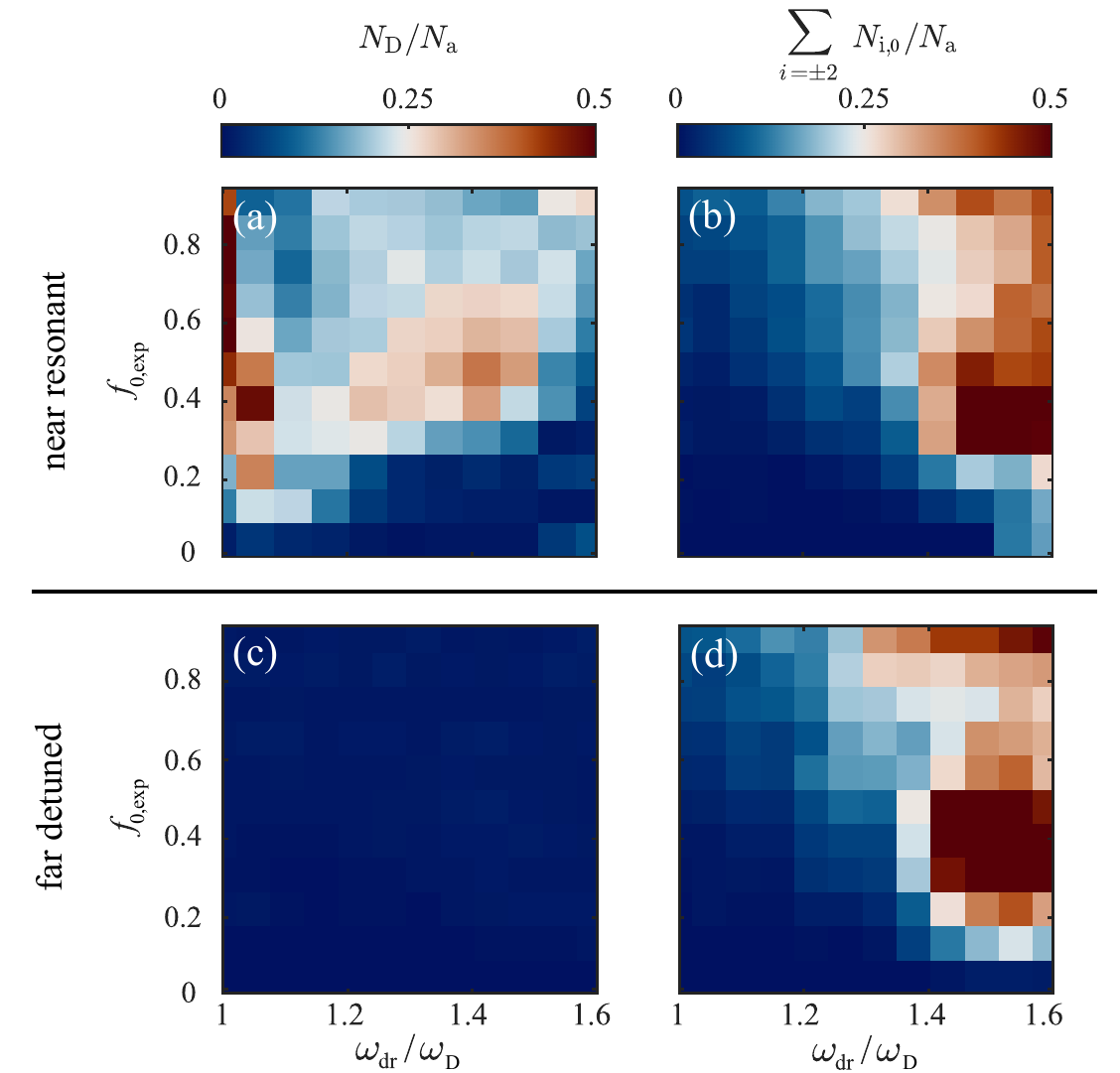}
\caption{Relative population of the dark state for different driving frequencies $\omega_\mathrm{dr} / \omega_\mathrm{D}$ and driving strengths $f_\mathrm{0}$ for (a) the pump light close to  resonance ($\delta_\mathrm{C}=~-2\pi \times 30$~kHz) and (c) far detuned from cavity resonance after six driving cycles. The driving frequency axis is rescaled by the resonance frequency of the dark state $\omega_\mathrm{D}$. (b) and (d) show the relative population of the $\{p_\mathrm{y}, p_\mathrm{z}\}=\{\pm 2, 0\}\hbar k$ momentum modes, which indicates the atoms populating the third band.}
\label{sf:3} 
\end{figure} 

We present in Fig.~\ref{sf:3}(a),(c) the experimentally obtained phase diagrams showing the population of the dark state with respect to the driving frequency $\omega_\mathrm{dr} / \omega_\mathrm{D}$ and driving strength $f_\mathrm{0}$. In Figs.~\ref{sf:3}(b),(d), we show the population of the $\{p_\mathrm{y}, p_\mathrm{z}\}=\{\pm 2, 0\}\hbar k$ momentum modes for the pump light, close and far detuned with respect to the cavity resonance. For the far detuned case, the cavity is basically inactive and we do not observe population of the dark state, which demonstrates the importance of cavity photons for the excitation of the dark state. Moreover, the parameter range, wherein we observe population of the $\{p_\mathrm{y}, p_\mathrm{z}\}=\{\pm 2, 0\}\hbar k$ momentum modes, is very similar for both cases, only its amplitude increases for the far detuned case since there are no atoms pumped into the dark state. \\
\indent As explained in the second paragraph of Sec.I, the dark state frequency is larger in the experiment than in our theoretical models. In our experimental observations, the resonance frequency for excitation of the $\{p_\mathrm{y}, p_\mathrm{z}\}=\{\pm 2, 0\}\hbar k$ momentum modes lies at $\approx 1.45 \, \omega_\mathrm{dr}/\omega_\mathrm{D}$. In SFig.~\ref{sf:3}(a), we see that in fact the transfer of atoms into the dark state is suppressed if the $\{\pm 2, 0\}\hbar k$ resonance is approached. Rather, the atoms are transferred into the second band, as shown in SFig.~\ref{sf:3}(b), without photon scattering into the cavity. The plots in (c) and (d) for large pump-cavity detuning $\delta_\mathrm{C}$ show that no dark state population arises, while the direct excitation of the $\{\pm 2, 0\}\hbar k$ momentum modes prevails. Finally, in our numerical models, the $\{\pm 2, 0\}\hbar k$ resonance arises at $\approx 1.7 \omega_\mathrm{dr}/\omega_\mathrm{D}$, which lies outside of the range of our simulations. 

\section{Atom-cavity system}
Our system is well described by the Hamiltonian \cite{Ritsch2013, Cosme2019, Skulte2021, Kongkhambut2021} 
\begin{align}
\label{eq:ham}
\hat{H}&/\hbar = - \delta_{\mathrm{C}} \hat{a}^\dagger \hat{a} + \int dy dz \hat{\Psi}^\dagger (y,z) \biggl[ -\frac{\hbar}{2m} \nabla^2 -\omega_\mathrm{rec}\epsilon_\mathrm{p} \cos^2(ky+\phi(t)) + U_0 \, \hat{a}^\dagger \hat{a} \cos^2(kz)  \\ \nonumber
&-\sqrt{\omega_\mathrm{rec}|U_0|\epsilon_\mathrm{p}}\cos(ky+\phi(t))\cos(kz)(a^\dagger+a)  \biggr] \hat{\Psi}(y,z),
\end{align}
where $\delta_\mathrm{C}$ is the pump-cavity detuning, and $U_0 < 0$ is the frequency shift of the cavity resonance due to a single atom ($|U_0| = 2 \pi \times 0.4 \,$kHz). The phase of the pump field is periodically driven according to $\phi(t)=f_0 \sin(\omega_\mathrm{dr}t)$ with the modulation index $f_0$ and the modulation frequency $\omega_\mathrm{dr}$. Furthermore, $\hat{a}~(\hat{a}^\dagger)$ is the annihilation (creation) operator for a photon in the single-mode cavity, while $ \hat{\Psi}~(\hat{\Psi}^\dagger)$ is the bosonic annihilation (creation) field operator for the atoms. Here, $k$ denotes the wave number of the pump light, $\epsilon_\mathrm{p}$ is the the pump strength, quantified in terms of the maximal energy depth of the pump lattice in units of the recoil energy $E_\mathrm{rec} = \hbar \omega_\mathrm{rec}$ with the recoil frequency $\omega_\mathrm{rec}=\hbar k^2/2m$, where $m$ is the atomic mass. The experiment operates in the recoil-resolved regime, i.e., the loss rate of the cavity photons $\kappa$ is smaller than the recoil frequency $\omega_\mathrm{rec}$. For our system $\omega_\mathrm{rec}= 2 \pi \times 3.6~\mathrm{kHz}$. We emphasize the importance of the recoil-resolved regime \cite{Kessler2014, Klinder2016} to excite the atoms into the dark state, as the underlying mechanism relies on a coherent coupling of a limited number of momentum modes.

\section{Processes for populating the dark state}
We briefly discuss the different scattering channels for populating the dark state in our theoretical models, i.e., the three-mode Dicke model and full-atom cavity simulations, and in the experimental setup. \\
\indent First, we discuss the difference between the full atom-cavity model and the three-level Dicke model. In the full atom-cavity system, we achieve a higher dark state population as compared to the results of the three-mode Dicke model after 7 driving cycles. In the three-mode model, we only consider momentum modes up to $\{p_\mathrm{y}, p_\mathrm{z}\}= \{\pm 1, \pm 1\}\hbar k$ and neglect the $\{p_\mathrm{y}, p_\mathrm{z}\}=\{\pm 2, 0\}\hbar k$ modes. However, as can be seen from the last line in the Hamiltonian in Eq.~\ref{eq:ham4}, atoms in the $\{p_\mathrm{y}, p_\mathrm{z}\}=\{\pm 2, 0\}\hbar k$ modes can be transferred into the dark state $|\mathrm{D}\rangle \equiv$ $\sum_{\nu,\mu\in\{-1,1\}} \nu |\nu \hbar k,\mu \hbar k\rangle$. This enhances the dark state population in the full atom-cavity system as compared to the three-mode Dicke model. \\
\indent Next, we discuss the population of the dark state in the experiment. While the experiment includes the channel for scattering from $\{p_\mathrm{y}, p_\mathrm{z}\}=\{\pm 2, 0\}\hbar k$ into the dark state, there are additional factors that decrease the efficiency of populating the dark state, i.e., heating and atom loss introduced by phase modulation of the pump wave. The atom loss effectively shifts the critical pump strength required to enter the superradiant phase, as the number of scatterers of photons decreases. Since cavity-photon-mediated interactions are necessary for the transfer of atoms into the dark state, atom loss, which decreases the occupation of the cavity mode, attenuates the process of populating the dark state. Furthermore, in the experiment, as discussed in Sec.I, the additional side lobes of the pump beam frequency spectrum push the dark state resonance towards higher frequencies. This effectively reduces the regime where the dark state can be populated, which is restricted to driving frequencies smaller than the resonance for excitation of the $\{p_\mathrm{y}, p_\mathrm{z}\}=\{\pm 2, 0\}\hbar k$ momentum modes. The latter resonance gives rise to an efficient transfer of the atoms into the maximum of the second band of the pump wave potential. The relevant driving term in Eq.~\ref{eq:ham} is $2 \phi(t) \sin(2ky)$, which arises by approximating $\cos(ky+\phi(t))^2$ for the case of small driving strengths. We note that the corresponding resonance frequency is light-shifted by the pump beam, however, this effect can be neglected for the relatively shallow pump lattice used in this work.

\section{Three-level system}
As first shown in \cite{Skulte2021}, the Hamiltonian in Eq.~\ref{eq:ham} can be mapped onto a parametrically driven dissipative three level model. Here, to capture the effects for strong driving, where $f_0 \ll 1$ is not fulfilled, we use trigonometric identities and the following Jacobi-Anger expansions \cite{abramowitz}
\begin{align}
\cos(z \sin(\theta)) &= J_0(z)+2 \sum_{n=1}^\infty J_{2n}(z) \cos(2n\theta) \\
\sin(z \sin(\theta)) &= 2 \sum_{n=1}^\infty J_{2n-1}(z) \cos((2n-1)\theta).
\end{align}
The Hamiltonian in Eq.~\ref{eq:ham} acquires the form
\begin{align}
\label{eq:ham4}
\hat{H}&/\hbar = - \delta_{\mathrm{C}} \hat{a}^\dagger \hat{a} + U_0  \hat{a}^\dagger \hat{a}  \int dy dz  \hat{\Psi}^\dagger (y,z)\cos^2(kz)\hat{\Psi} (y,z) \\ \nonumber
&-\omega_\mathrm{rec}\epsilon_\mathrm{p} \int dy dz  \hat{\Psi}^\dagger (y,z) \frac{1+\cos(2ky) \left[J_0(2f_0)+2h_2(t)\right]-2\sin(2ky) g_2(t)}{2} \hat{\Psi} (y,z)  \\  \nonumber
&-\sqrt{\omega_\mathrm{rec}|U_0|\epsilon_\mathrm{p}}(a^\dagger+a) \int dy dz  \hat{\Psi}^\dagger (y,z)\cos(ky)\cos(kz)\left(J_0(f_0)+2h_1(t)\right) \hat{\Psi} (y,z) \\ \nonumber
&+\sqrt{\omega_\mathrm{rec}|U_0|\epsilon_\mathrm{p}}(a^\dagger+a) \int dy dz  \hat{\Psi}^\dagger (y,z)\sin(ky)\cos(kz)2g_1(t) \hat{\Psi} (y,z),
\end{align}
where we defined  $h_2(t)=\sum_{n=1}^\infty J_{2n}(2f_0)\cos(2n\omega_\mathrm{dr} t)$ and $g_2(t)=\sum_{n=1}^\infty J_{2n-1}(2f_0)\sin((2n-1)\omega_\mathrm{dr} t)$ and  $h_1(t)=\sum_{n=1}^\infty J_{2n}(f_0)\cos(2n\omega_\mathrm{dr} t)$ and $g_1(t)=\sum_{n=1}^\infty J_{2n-1}(f_0)\sin((2n-1)\omega_\mathrm{dr} t) $. \\
Next, the atomic field operator is approximated as
\begin{equation}
\hat{\Psi}(y,z)=\hat{c}_0 \psi_0(y,z)+ \hat{c}_1 \psi_1(y,z)+ \hat{c}_2 \psi_2(y,z)
\end{equation}
where $\hat{c}_i$ are bosonic annihilation operator, and $\psi_0(y,z)=1$, $\psi_1(y,z)=2 \cos(ky)\cos(kz)$ and $\psi_2(y,z)=2\sin(ky)\cos(kz)$. \\ We note that in applying this approximation we neglect higher momentum mode contributions, e.g. $\cos(2ky)$, which contribute heavily for higher driving frequencies around $\sim 14.5~\mathrm{kHz}$ as can be seen in Fig.~\ref{sf:3}.
\\ Under parity change $y\rightarrow -y$ these wave functions transform as 
\begin{align}
\mathcal{P}_y \psi_0(y,z)&=\psi_0(-y,z)=+\psi_0(y,z) \\
\mathcal{P}_y \psi_1(y,z)&=\psi_1(-y,z)=+\psi_1(y,z) \\
\mathcal{P}_y \psi_2(y,z)&=\psi_2(-y,z)=-\psi_2(y,z)~.
\end{align}
Hence, only $\psi_2$ gets a minus sign upon application of $\mathcal{P}_y$. Using a Schwinger boson representation, the bosonic operators can be mapped onto pseudo-spin operators to obtain a driven three-level Dicke Hamiltonian
\begin{align}\label{eq:ham2}
H/ \hbar &= \omega \hat{a}^\dagger \hat{a} +\left(\omega_{\mathrm{B}}-\Delta_{f_0}\right)\hat{J}^{\mathrm{B}}_z   + \left(\omega_{\mathrm{D}}+\Delta_{f_0}\right)\hat{J}^{\mathrm{D}}_z  +f_2(t)\left(\omega_{\mathrm{B}}-\omega_{\mathrm{D}} \right) \left(\hat{J}^{\mathrm{D}}_z -\hat{J}^{\mathrm{B}}_z  \right)+2g_2(t) \left(\omega_{\mathrm{B}}-\omega_{\mathrm{D}} \right) \hat{J}^{\mathrm{BD}}_x
 \\ \nonumber &+\frac{2\left(\lambda_{f_0}+\eta(t)\right)}{\sqrt{N}}\left(\hat{a}^\dagger+\hat{a} \right)  \hat{J}^{\mathrm{B}}_x
 -\frac{2 \zeta(t)}{\sqrt{N}}\left(\hat{a}^\dagger+\hat{a} \right)\hat{J}^{\mathrm{D}}_x,
\end{align}
where $\omega_{\mathrm{D}}=2 \omega_\mathrm{rec}(1-\frac{\epsilon_\mathrm{p}}{8})$, $\omega_{\mathrm{B}}=2 \omega_\mathrm{rec}(1+\frac{\epsilon_\mathrm{p}}{8})$, $\Delta_{f_0}=\frac{\epsilon_\mathrm{p}\omega_\mathrm{rec}}{4}(1-J_0(2f_0))$, $2  \lambda \equiv \sqrt{N_{\mathrm{a}} \epsilon_\mathrm{p} \omega_\mathrm{rec} |U_0|}$, $\lambda_{f_0}=J_0(f_0)\lambda$, $\eta(t)=2h_1(t)\lambda$ and $\zeta(t)=2g_1(t)\lambda$. Expanding this Hamiltonian up to linear order in the driving strength $f_0$ leads to the parametrically driven dissipative three-level Dicke model presented in \cite{Skulte2021,Kongkhambut2021}.

\begin{figure}[!htpb]
\centering
\includegraphics[width=0.5\columnwidth]{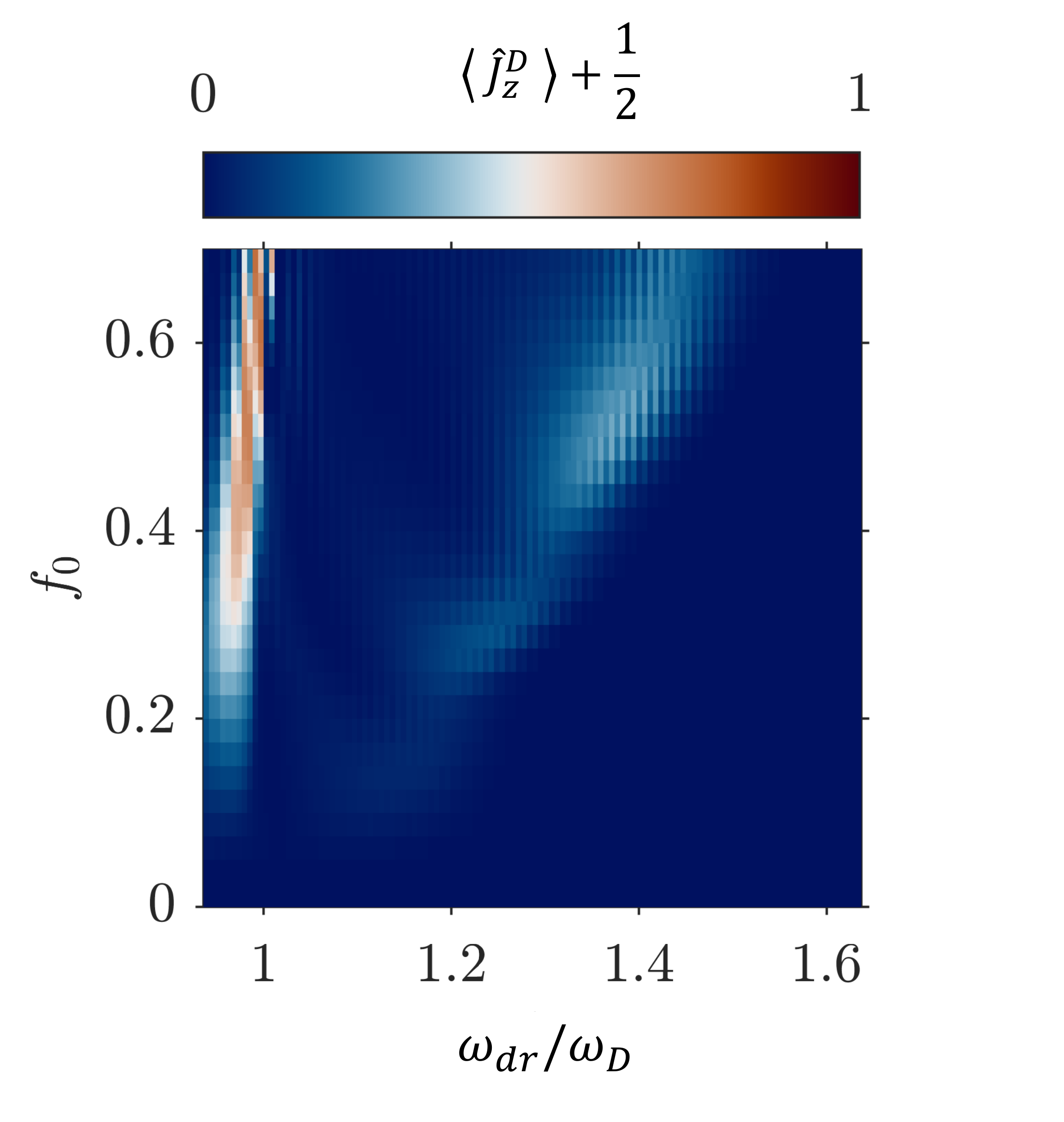}
\caption{ Population of the dark state for different driving frequencies $\omega_\mathrm{dr}$ and driving strengths $f_{0}$ calculated for the Hamiltonian Eq.~\ref{eq:ham3}. The driving frequency axis is rescaled by the characteristic frequency of the dark state $\omega_\mathrm{D}$. The phase diagram is constructed for 7 driving cycles.}
\label{sf:4} 
\end{figure} 

\subsection{Large $\kappa$ limit}
Within our three-level Dicke model we can adiabatically eliminate the light field, if $\kappa \gg \omega_{\textrm{rec}}$. That is, we assume $\frac{d a}{d t} \approx 0$ and solve for $a$ to obtain an atom-only or spin-only like three-level model
\begin{align}\label{eq:ham3}
\hat{H}_\mathrm{eff}/\hbar &=\left(\omega_{\mathrm{B}}-\Delta_{f_0}\right)\hat{J}^{\mathrm{B}}_z   + \left(\omega_{\mathrm{D}}+\Delta_{f_0}\right)\hat{J}^{\mathrm{D}}_z  +f_2(t)\left(\omega_{\mathrm{B}}-\omega_{\mathrm{D}} \right) \left(\hat{J}^{\mathrm{D}}_z -\hat{J}^{\mathrm{B}}_z  \right)+2g_2(t) \left(\omega_{\mathrm{B}}-\omega_{\mathrm{D}} \right) \hat{J}^{\mathrm{BD}}_x
 \\ \nonumber & - \Lambda \left( \left(J_0(f_0)+2 h_1(t)\right)^2 \hat{J}^{\mathrm{D}}_x \hat{J}^{\mathrm{D}}_x -\left(J_0(f_0)+2 h_1(t)\right)2g_1(t)\left[\hat{J}^{\mathrm{D}}_x \hat{J}^{\mathrm{B}}_x+\hat{J}^{\mathrm{B}}_x \hat{J}^{\mathrm{D}}_x \right] + 4g_1(t)^2\hat{J}^{\mathrm{B}}_x \hat{J}^{\mathrm{B}}_x\right),
\end{align}
with $\Lambda=8\lambda^2 \omega /(N(\kappa^2+\omega^2))$. This is the three-level generalisation of the prescription for mapping the standard two-level Dicke model onto the Lipkin-Meshkov-Glick model by adiabatically eliminating the photon dynamics \cite{Keeling2010}. In Fig.~\ref{sf:4}, the corresponding phase diagram for varying driving strength and driving frequency is shown for $\lambda= 1.05\, \lambda_\mathrm{crit}$. Note, that in a full description of the atom-cavity setup in terms of Eq.~\ref{eq:ham}, a large value of $\kappa$ would enable the excitation of higher modes, not included here, with the consequence of decoherence and heating.

\section{Dark state condensation below the critical pump strength}

\begin{figure}[!htbp]
\centering
\includegraphics[width=0.5\columnwidth]{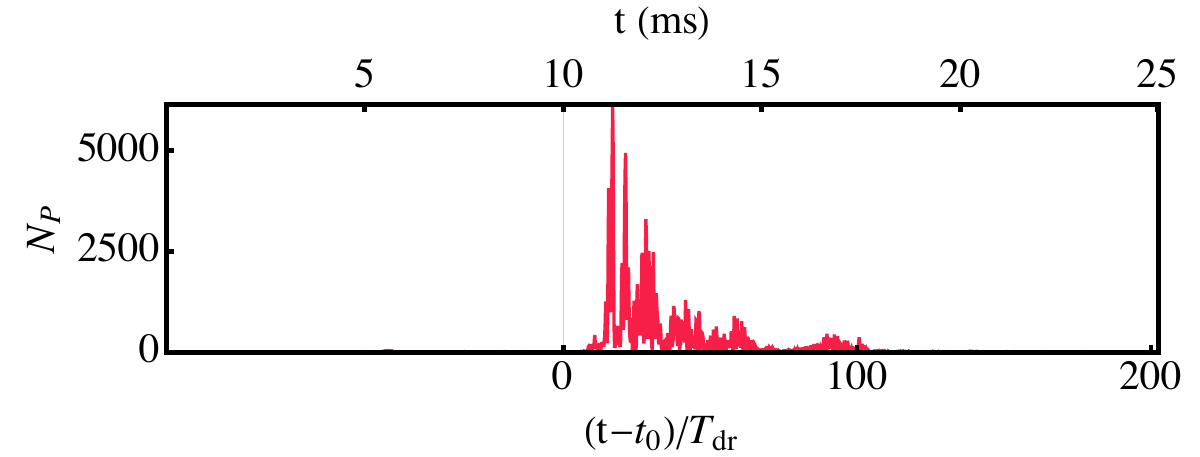}
\caption{Number of photons inside the cavity. The periodic drive is switched on at $t_0$. The parameters are the same as those used in Fig. 4 except for $\epsilon \approx 0.96\,\epsilon_\mathrm{crit}$.}
\label{sf:5} 
\end{figure}

\begin{figure}[!htbp]
\centering
\includegraphics[width=0.45\columnwidth]{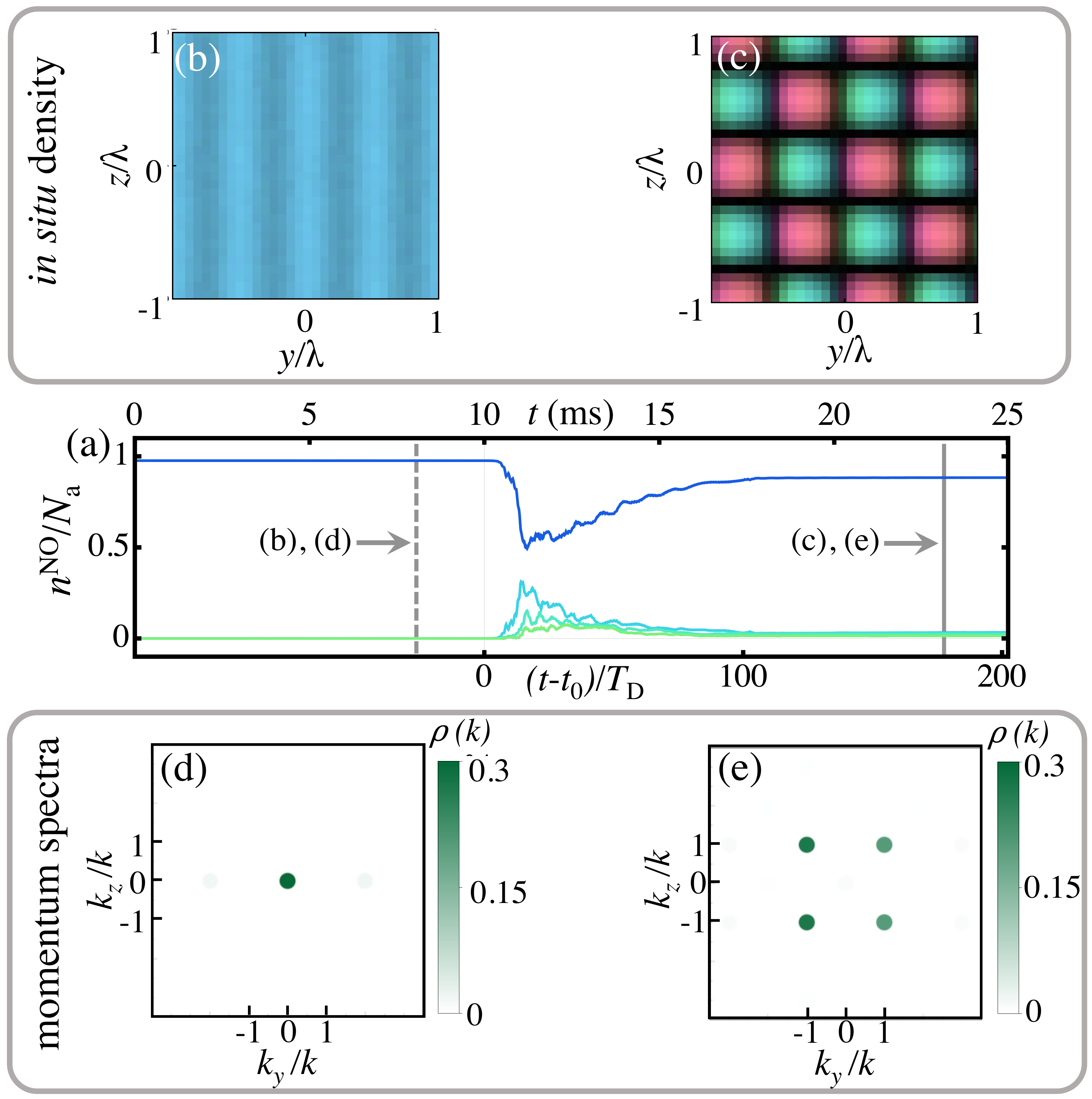}
\caption{(a) TWA results for the evolution of the five highest eigenvalues of the single-particle correlation function at equal time (SPCF). Gray vertical lines denote the times when the snapshots of the single-particle density in (b) and (c) are taken. (b), (c) The real space densities are color coded to show the phase within the $(y,z)$-plane. (d), (e) Momentum spectra at times indicated in (a).}
\label{sf:6} 
\end{figure}

Here, we briefly show the dark state condensation starting below the critical pump strength. We ramp up the pump strength to $\epsilon\approx 0.96\, \epsilon_\mathrm{crit}$ and start the modulation after $10~\mathrm{ms}$. In Fig.~\ref{sf:5}, it can be seen that after the modulation is switched on, the light field builds up before it vanishes again after a large fraction of atoms occupies the dark state as can be seen from the long-time behaviour in Fig.~\ref{sf:6}. This again highlights the importance of the intra-cavity field for transfering the atoms into the dark state. We note that the transition into the dark state is slower compared to the case starting from the superradiant phase discussed in the main text.

\section{Mode population during the ramp-down process}
\begin{figure}[!htbp]
\centering
\includegraphics[width=0.8\columnwidth]{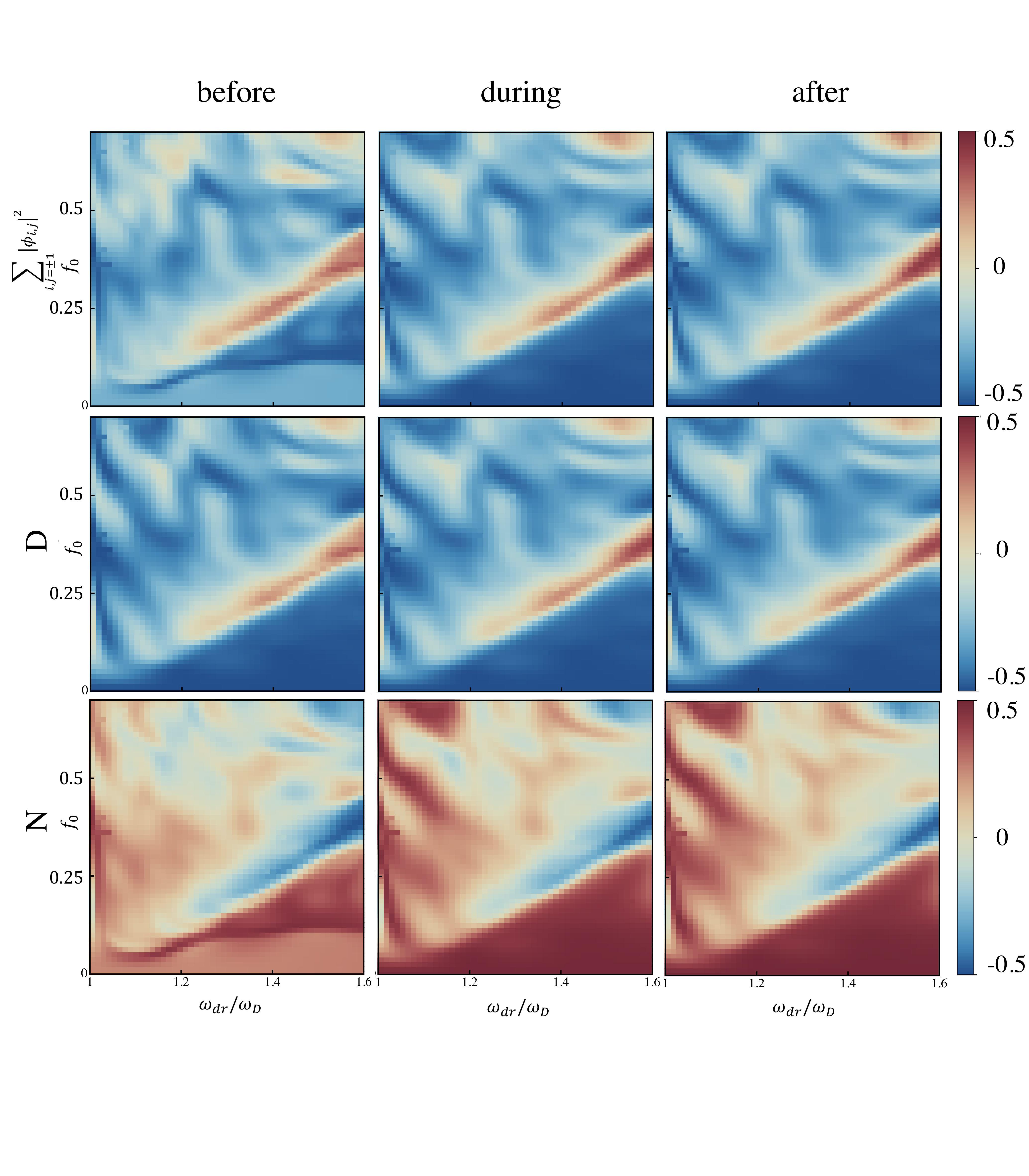}
\caption{Population of the $\{\pm1,\pm1\}\hbar k$ momentum modes, the dark state and the ground state for different driving frequencies $\omega_\mathrm{dr}$ and driving strengths $f_0$. The driving frequency axis is rescaled by the characteristic frequency of the dark state, $\omega_\mathrm{D}$.}
\label{sf:7} 
\end{figure}

Fig.~\ref{sf:7} presents the occupation of the sum of the $\{\pm1,\pm1\}\hbar k$ momentum modes, the $|\mathrm{D}\rangle$ as well as the $|\mathrm{N}\rangle$, before, during and after the ramp-down of the pump laser for varying driving strength and driving frequencies rescaled by the characteristic dark state frequency $\omega_\mathrm{D}$. Before the ramp-down process the population of the $|\mathrm{D}\rangle$ and $|\mathrm{B}\rangle$ cannot be distinguished by summing up the $\{\pm1,\pm1\}\hbar k$ momentum modes in a TOF image. However during the ramp-down the populations of $|\mathrm{B}\rangle$ is transferred back into the $|\mathrm{N}\rangle$ and the phase diagram of the population of the $\{\pm1,\pm1\}\hbar k$ momentum modes and the population of the dark state $|\mathrm{D}\rangle$ are approximately the same. This motivates us to measure the population of $|\mathrm{D}\rangle$ using this scheme.

\black
\bibliography{references_dark_state}